\documentstyle [aps,prd]{revtex}
\input epsf
\begin{document}

\newcommand{\alp}{$\alpha\,\,$}
\newcommand{\xe}{$x_e\;$}
\newcommand{\tdot}{$\dot{\tau}\;\,$}
\newcommand{\be}{\begin{equation}}
\newcommand{\ee}{\end{equation}}
\newcommand{\obh}{$\Omega_B h^2\;$}
\newcommand{\omh}{$\Omega_m h^2\;$}
\newcommand{\och}{$\Omega_c h^2\;$}		
\newcommand{\okh}{$\Omega_K h^2\;$}	
\newcommand{\olh}{$\Omega_\Lambda h^2\;$}
\newcommand{\deln}{$\Delta N_\nu\;$}
\newcommand\la{\lower0.6ex\vbox{\hbox{$ \buildrel{\textstyle 
<}\over{\sim}\ $}}}
\newcommand\ga{\lower0.6ex\vbox{\hbox{$ \buildrel{\textstyle 
>}\over{\sim}\ $}}}
	
\title{How Does CMB + BBN Constrain New Physics?}
\author{James P. Kneller$^{1}$, Robert J. Scherrer$^{1,2}$, 
Gary Steigman$^{1,2}$, and Terry P. Walker$^{1,2}$}
\address{$^1$Department of Physics, The Ohio State University,
Columbus, OH~~43210}
\address{$^2$Department of Astronomy, The Ohio State University,
Columbus, OH~~43210}
\date{\today}
\maketitle







\begin{abstract}

Recent cosmic microwave background (CMB) results from BOOMERANG, 
MAXIMA, and DASI provide cosmological constraints on new physics 
that can be competitive with those derived from Big Bang Nucleosynthesis 
(BBN).  In particular, both CMB and BBN can be used to place limits 
on models involving neutrino degeneracy and additional relativistic 
degrees of freedom.  However, for the case of the CMB, these 
constraints are, in general, sensitive to the assumed priors.  
We examine the CMB and BBN constraints on such models and study 
the sensitivity of ``new physics" to the assumed priors.  If we 
add a constraint on the age of the universe ($t_0 ~\ga 11$~Gyr), 
then for models with a cosmological constant, the range of baryon 
densities and neutrino degeneracy parameters allowed by the CMB 
and BBN is fairly robust: $\eta_{10} = 6.0 \pm 0.6$, \deln $\la 6$, 
$\xi_e ~\la 0.3$.  In the absence of new physics, models without a 
cosmological constant are only marginally compatible with recent CMB 
observations (excluded at the 93\% confidence level).

\end{abstract}

\twocolumn

\section{INTRODUCTION}

Until recently, Big Bang Nucleosynthesis (BBN) provided the only precision 
estimates of the baryon density of the universe.  Based on recent deuterium 
observations \cite{BT}, BBN identifies a value for the baryon density which 
has been variously estimated (depending on the choice for the primordial 
deuterium abundance) as $\Omega_B h^2 = 0.015 - 0.023$ \cite{OSW}, and 
$\Omega_B h^2 = 0.017 - 0.021$ \cite{BNT}, or incorporating the most recent 
data $\Omega_B h^2 = 0.017 - 0.024$ \cite{omeara}, where $\Omega_B$ is the 
baryon density expressed as a fraction of the critical density, $h$ is the 
Hubble parameter in units of 100 km/sec/Mpc, and the ranges quoted are 
intended to be at the 95\% confidence level.

In the past year, observations of the cosmic microwave background (CMB) 
fluctuations
have become a competitive means for estimating the baryon density.  
These data have been used both alone and in combination with other 
observations (such as type Ia supernovae and large-scale structure) 
to set limits on $\Omega_B h^2$.  The preliminary CMB data from BOOMERANG 
\cite{boom} and MAXIMA \cite{max} suggested a higher baryon density 
($\Omega_B h^2 \sim 0.03$) than that predicted from BBN, due to the 
unexpectedly low second acoustic peak in these CMB observations (see, 
for example, Ref. \cite{Lange} - \cite{white}).  This discrepancy has 
vanished in the wake of more recent data from BOOMERANG \cite{boom2}, 
MAXIMA \cite{max2} and DASI \cite{dasi}.

This original discrepancy between the BBN and CMB predictions for 
$\Omega_B h^2$ led to the suggestion that perhaps new physics must 
be invoked to reconcile the BBN and CMB predictions for $\Omega_B h^2$.  
The problems for BBN at the high baryon density suggested by the Refs.
\cite{Lange} - \cite{white} are that the BBN-predicted abundance of 
deuterium is too low while those of helium-4 and lithium-7 are too 
high when compared to the observationally inferred primordial abundances.  
If, however, the universal expansion rate were increased during the BBN 
epoch by, for example, the contribution to the total energy density of 
``new" neutrinos and/or other relativistic particles, the BBN-predicted 
abundance of deuterium would increase (less time for D-destruction), 
while that of lithium would decrease (less time for production of 
$^7$Be).  This increase in the expansion rate results in a higher 
helium abundance, but the BBN-predicted helium abundance can be 
reduced by a non-zero chemical potential for the electron neutrinos.  
An excess of $\nu_{e}$ over $\bar{\nu}_{e}$ can drive the neutron-proton 
ratio down, leading to reduced production of helium-4.  Thus, reconciling 
BBN with a high baryon density would require two kinds of ``new physics'': 
the expansion rate should be faster than the standard value and $\nu_{e}$ 
should be ``degenerate''.  Although these two effects may be unrelated,
neutrino degeneracy can provide an economic mechanism for both, since
the energy density contributed by degenerate neutrinos exceeds that
from non-degenerate neutrinos, leading to an enhanced expansion rate
during the epoch of BBN. As Kang \& Steigman \cite{ks} and Olive et al.
\cite{ostw} have shown, the observed primordial abundances of the light
nuclides can be reconciled with very large baryon densities provided
sufficient neutrino degeneracy is permitted.  

Although the most recent CMB observations suggest that no new physics 
need be invoked to reconcile the CMB and BBN observations, these 
measurements also provide another tool, independent of BBN, to constrain 
such new physics.  From the contribution of Ref. \cite{lesg} and the 
combined CMB and BBN analyses of Ref. \cite{esp2} -\cite{hann1}, it 
is already clear that the constraints on ``new physics" are strongly 
dependent on the priors assumed in the analysis for the other, non-BBN 
related cosmological parameters.  Here we explore this issue further.  
In particular, we consider the concordance  between the CMB and BBN 
predictions for $\Omega_B h^2$ in models with neutrino degeneracy using 
four different representative sets of priors.  In the next section we 
discuss our calculation and give results for our four models.  Our 
conclusions are summarized in Secs.~3 and ~4.

\section{CALCULATIONS}

Our first step is the calculation of element abundances in BBN for models 
with degenerate neutrinos.  This is a well-understood calculation with a 
long history, and the reader is referred to Refs. \cite{ks}, \cite{ostw}, 
\cite{degen} for the details.  

The degeneracy of any of the three neutrinos increases the total relativistic 
energy density, leading to an increase in the overall expansion rate.  
During ``radiation dominated'' epochs the expansion rate (Hubble parameter) 
is proportional to the square root of the total energy density in extremely 
relativistic (ER) particles so the speedup factor, $S$, is
\begin{equation}
S \equiv H'/H = (\rho '/\rho)^{1/2}.
\end{equation}

In addition, the {\sl electron} neutrino separately affects the rates 
of the weak reactions which interconvert 
protons and neutrons, and so
it is convenient to parameterize
the neutrino degeneracy in terms of $\xi_e$ and $\Delta N_\nu$, where
$\xi_e = \mu_e/T_\nu$ is the ratio of the electron neutrino chemical 
potential $\mu_e$ to the neutrino temperature $T_\nu$, and $\Delta N_\nu$ 
($\equiv N_{\nu} - 3$) is the additional energy density contributed by 
all the degenerate neutrinos {\it as well as any other energy density 
not accounted for in the standard model of particle physics} (e.g., 
additional relativistic particles) expressed in terms of the equivalent 
number of extra, non-degenerate, two-component neutrinos:
\begin{equation}
\rho' - \rho \equiv \Delta \rho_{ER} \equiv \Delta N_{\nu}
\rho_{\nu}(\xi = 0).
\end{equation}
The contribution to \deln from one species of degenerate neutrinos 
is \cite{ks},
\begin{equation}
\Delta N_{\nu} = 15/7[(\xi/\pi)^4 + 2(\xi/\pi)^2],
\end{equation}
We emphasize that our results are independent of whether \deln 
(or, equivalently, the corresponding value of $S$) arises from neutrino 
degeneracy, from ``new'' (ER) particles, or from some other source.
Note that a non-zero value of $\xi_e$ implies a non-zero contribution to 
$\Delta N_\nu$ from the electron neutrinos alone; we have 
included this contribution in our calculations. However, for the range 
of $\xi_e$ which proves to be of interest for BBN consistency ($\xi_e 
~\la 0.5$), the degenerate electron neutrinos contribute only a small 
fraction of an additional neutrino species to the energy density 
($\Delta N_\nu ~\la 0.1$).

The question we address is: for a given value of the baryon-photon 
ratio $\eta$ ($\eta_{10} \equiv 10^{10}\eta = 274\Omega_{B}h^{2}$), 
are there values for $\xi_e$ and for $\Delta N_\nu$ which result in 
agreement between the BBN predictions and the known limits on the 
primordial element abundances?  Through the hard work of many observers, 
aided by better detectors and bigger telescopes, the statistical 
uncertainties in the observationally inferred primordial abundances 
have been reduced significantly in recent years.  In contrast, the 
systematic errors are still quite large (cf. \cite{OSW}).  For this 
reason we adopt generous ranges for the primordial abundances of $^4$He, 
D, and $^7$Li.  Furthermore, even for fixed values of $\eta$, $\xi_e$, 
and $\Delta N_\nu$, there are uncertainties in the BBN-predicted abundances 
due to uncertainties in the nuclear and/or weak reaction rates.  We have 
chosen the ranges for the primordial abundances large enough to encompass 
these uncertainties as well.  For the primordial helium-4 mass fraction, 
we take the limits to be
\begin{equation}
\label{BBN1}
0.23 \le {\rm Y}_P \le 0.25.
\end{equation}
For deuterium and lithium-7, expressed as number ratios to hydrogen,
we take the limits
\begin{equation}
\label{BBN2}
2 \times 10^{-5} \le {\rm D/H} \le 5 \times 10^{-5},
\end{equation}
and
\begin{equation}
\label{BBN3}
1 \times 10^{-10} \le {\rm ^7Li/H} \le 4 \times 10^{-10}.
\end{equation}

Our allowed parameter range is thus a three-dimensional volume
in the space of $\eta$, $\xi_e$, and $\Delta N_\nu$.  However,
since we wish to compare to the predictions of the CMB, which are
sensitive to $\eta$ and $\Delta N_\nu$, but independent of $\xi_e$,
we project our allowed BBN region onto the $\eta - \Delta N_\nu$ 
plane.  Our BBN results are shown in the four panels of Figure 
1 where, for four choices of $\xi_e$ we show the iso-abundance 
contours for Y$_{P}$, D/H and Li/H in the $\eta - \Delta N_\nu$ 
plane.  The shaded areas highlight the acceptable regions in our 
parameter space.  As $\xi_e$ increases, the allowed region moves 
to higher values of $\eta$ and $\Delta N_\nu$, tracing out a 
BBN-consistent band in the $\eta - \Delta N_\nu$ plane.  This band 
is shown by the dashed lines in Figures 2 \& 5.  The trends are easy 
to understand (see \cite{ks} \& \cite{ostw}).  As the baryon density 
increases the universal expansion rate (as measured by $\Delta N_\nu$) 
must increase to keep the deuterium and lithium unchanged, while 
the $\nu_{e}$ degeneracy ($\xi_{e}$) must increase to maintain the 
helium abundance at its SBBN value. 

\begin{figure}[ht]
\centering
\epsfysize=4.05truein 
\epsfbox{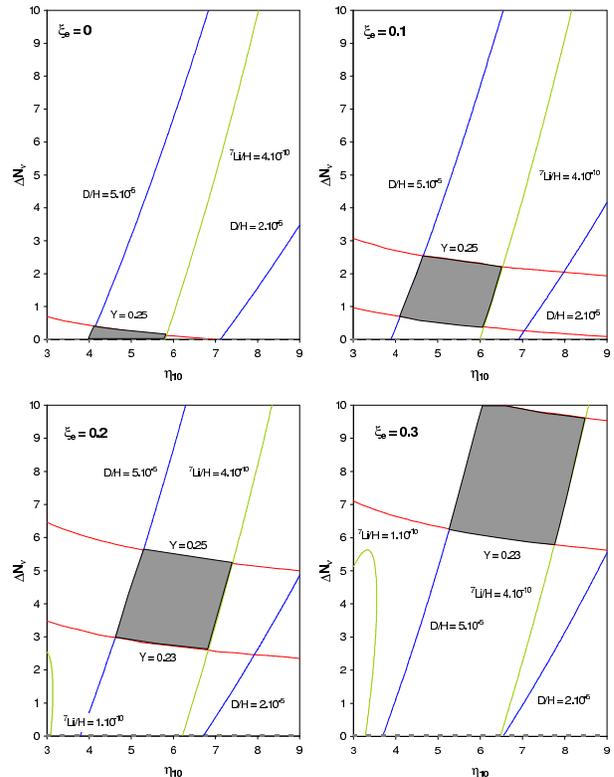}
	\caption{\small{Iso-abundance contours for deuterium 
(D/H), lithium (Li/H) and helium (mass fraction, Y) in the 
\deln -- $\eta_{10}$ plane for four choices of $\nu_{e}$ 
degeneracy ($\xi_{e}$).  The shaded areas highlight the range
of parameters consistent with the adopted abundance ranges 
(see eq. 10 -- 12).}}
 	\label{fig1}
\end{figure}

We then use CMBFAST \cite{CMBFAST} to calculate the CMB fluctuation 
spectrum as a function of $\eta$ and $\Delta N_\nu$ and compare with 
the BOOMERANG \cite{boom2} and MAXIMA \cite{max2} and DASI \cite{dasi} 
observations.  However, the CMB anisotropy spectrum is sensitive to a 
large number of other parameters which have no effect on BBN, including
the fraction of the critical density in non-relativistic matter $\Omega_M$ 
(where $\Omega_M$ includes both baryonic and non-baryonic matter), the 
fraction of the critical density contributed by the cosmological constant
$\Omega_\Lambda$ (or an equivalent vacuum energy density), the total 
$\Omega$ ($\equiv \Omega_M + \Omega_\Lambda$; $\Omega = 1$ corresponds 
to a ``flat'' universe), the Hubble parameter $h$, and the slope of the 
primordial power spectrum $n$ (``tilt").  Since we are interested in the 
way in which restricting these parameters affects the agreement between 
the CMB and BBN, we consider four representative sets of prior assumptions:

\vskip 0.4 cm

\noindent Case A:  
~$\Omega = 1$, ~$0.4 \le h \le 1.0$, ~$\Omega_B \le 
\Omega_M \le 1$,\\ ~$n=1$,

\noindent Case B:  
~$\Omega = 1$, ~$0.4 \le h \le 1.0$, ~$\Omega_B \le 
\Omega_M \le 1$, \\ ~$0.7 \le n \le 1.3$,

\noindent Case C:  
~$\Omega = 1$, ~$0.5 \le h \le 0.9$, ~$\Omega_B \le 
\Omega_M \le 0.4$, \\ ~$0.7 \le n \le 1.3$,

\noindent Case D:  
~$\Omega \le 1$, $0.5 \le h \le 0.9$, $\Omega_B \le 
\Omega_M = \Omega$, \\ $0.7 \le n \le 1.3$.

\vskip 0.4 cm

In models A -- C, the inflation-inspired assumption that the universe 
is flat is adopted and a cosmological constant is assumed to give 
$\Omega = 1$; in contrast, in model D the value of $\Omega_\Lambda$ 
is set to zero and the universe is allowed to be open or flat \cite{lesg}.  
Tensor modes are ignored in all of these cases.  Case A differs from 
Case B only in the restriction of tilt to $n = 1$.  Case C differs 
from Case B in the adoption of a slightly smaller range for the Hubble 
parameter and, of more significance, a more restricted range for the 
non-relativistic matter density, both of which are consistent with 
complementary observational data.  Case C is closest to what is often 
referred to as the ``concordance $\Lambda$CDM model" while Case D is 
closest to the ``SCDM model" which is inconsistent with the SN Ia data 
\cite{sn1a}.

\begin{figure}[ht]
	\centering
	\epsfysize=4.15truein 
\epsfbox{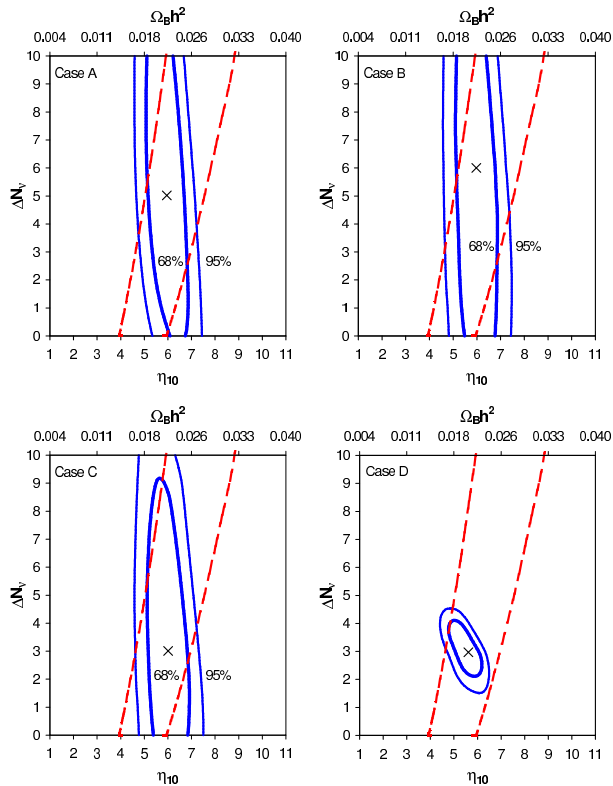}
	\caption{\small{68\% and 95\% contours (solid lines) for the 
	BOOMERANG, MAXIMA, and DASI CMB anisotropies in the \deln -- 
	$\eta_{10}$ plane (the upper horizontal axes show $\Omega_{B}h^{2}$) 
	for Cases A -- D.  The crosses indicate the best fit values.  
	The consistency band for BBN is shown by the dashed lines.}}
 	\label{fig2}
\end{figure}

For each of these sets of priors, we determine the best fit CMB model
for a given pair of values of $\Delta N_\nu$ and $\eta$ and assign 
a confidence limit based on the $\Delta \chi^2$ value calculated with
RADPACK \cite{bjk}.  In the four
panels of Figure 2 we display the (two-parameter) 68\% and 95\% contours 
in the $\eta$ -- $\Delta N_{\nu}$ plane for the four choices of priors 
discussed above.  The different shapes of the confidence 
interval contours highlight the sensitivity of the ``new physics'' 
($\Delta N_{\nu}$) to the choices of priors for the other cosmological 
parameters.

\section{DISCUSSION}

The effect on the post-BBN universe of a non-zero \deln is to enhance
the relativistic energy density, delaying the epoch of equal matter
and radiation densities.  This can be offset by increasing $\Omega_M$,
effectively restoring the original, \deln = 0, ratio of matter and 
radiation densities.  This effect produces the large difference between
cases A \& B and case C.  This may be seen in Figure 3 where the 
sensitivity of the constraints on $\Delta N_\nu$ to the priors adopted 
in the CMB fits is explored by comparing the $\chi^{2}$ distributions 
for our four Cases A -- D.  In cases A \& B, very large values for 
\deln are allowed, corresponding to large values of $\Omega_M$.  Thus, 
cases A \& B do not provide very effective upper limits on \deln when 
only the CMB data is taken into account (Fig.~3).  For case C, in 
contrast, large values of $\Omega_M$ are not permitted.  As seen in 
Figure 3 this results in a stronger upper bound on \deln: at the 68\% 
confidence level, \deln $< 6.7$.  Case D yields a very different set 
of constraints.  In this model, values of $\Omega_M < 1$ are compensated 
with curvature, rather than with a cosmological constant.  But the 
position of the first acoustic peak strongly constrains the curvature 
to be nearly zero, forcing $\Omega$ to be nearly unity.  Hence, in 
these models $\Omega_M \approx 1$, with almost no freedom to vary, 
and a change in \deln cannot be cancelled by changing $\Omega_M$.  
Thus, the allowed range for \deln is very small.

\begin{figure}[ht]
	\centering
	\epsfysize=4.0truein 
\epsfbox{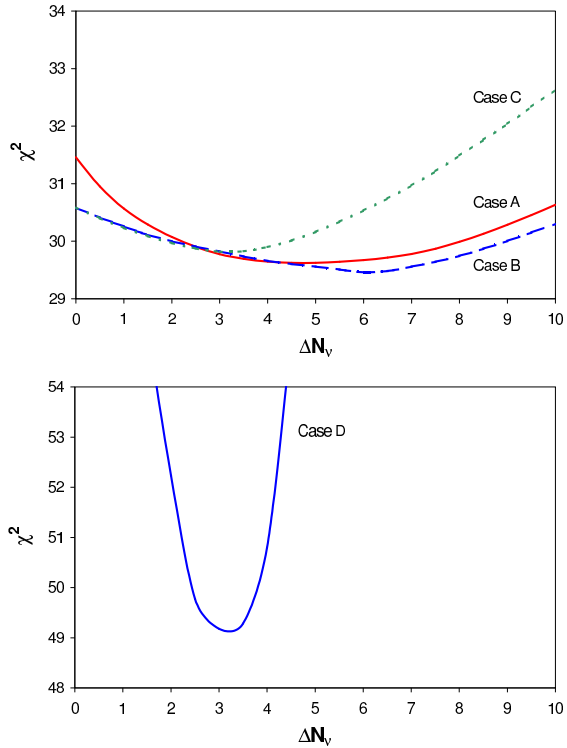}
	\caption{\small{$\chi^{2}$ distributions for \deln for the four 
	sets of priors corresponding to cases A -- D.}}
 	\label{fig3}
\end{figure}

Despite the differences, there are some striking similarities in the
parameter ranges identified in Figure 2.  With the exception of case D, 
the preferred ranges of baryon densities are very similar (see Figure 
4).  At 68\%  confidence $5.4~\la \eta_{10}~\la 6.6$ (at 95\% confidence, 
$4.8~\la \eta_{10}~\la 7.2$), for cases A -- C; for case D, $\eta_{10}$
is shifted downwards by $\approx 0.6$.  For all cases, a baryon density 
$\Omega_B h^2 \approx 0.02$ is a robust prediction of the CMB observations.

In contrast, constraints on the magnitude of the ``new physics'' 
($\Delta N_\nu$) do depend sensitively on the choice of priors.    
As noted earlier, for the $\Lambda$CDM models (cases A -- C), case C 
produces a stronger upper bound on $\Delta N_\nu$, than do cases A 
or B.  Figure 3 also illustrates a point which is only marginally 
apparent from Figure 2:  Case A prefers a non-zero value of \deln 
slightly more than do cases B and C (albeit not at a statistically 
significant level).  Since case A fixes $n=1$, this suggests that 
a nonzero \deln can mimic, to some extent the effect of ``tilt".  
This point is further emphasized when the BBN data are included in 
Figure 2: for \deln = 0, the overlap between the allowed values for 
$\eta$ for CMB and BBN is smaller for case A (ruled out at the 68\% 
confidence level), than for cases B and C.  However, given the 
marginal level of exclusion (68\%), this cannot be used to argue
for ``new physics".  In contrast, as already noted, case D is 
anomalous; in the absence of new physics it disagrees with the 
CMB data at the 93\% confidence level. 

\begin{figure}[ht]
	\centering
	\epsfysize=4.0truein 
\epsfbox{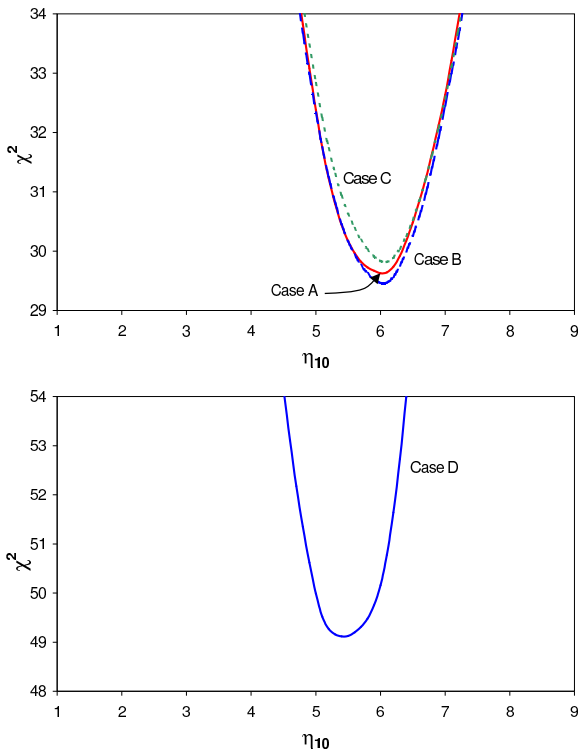}
	\caption{\small{$\chi^{2}$ distributions for $\eta_{10}$ for 
	the four sets of priors corresponding to cases A -- D.}}
 	\label{fig4}
\end{figure}

It is clear from Figure 3 that, with the exception of the statistically
disfavored case D, the CMB provides only very weak constraints on $\Delta
N_{\nu}$.  The notable constrast between cases B and C, with very similar 
priors, demonstrates the significant sensitivity of \deln to the choice 
of priors.  Because of this sensitivity, it is difficult to compare our 
results directly with those of Hannestad \cite{hann}, Lesgourgues \& Liddle 
\cite{LL}, and of Hansen et al. \cite{hansen}.  We are in agreement with 
Hannestad \cite{hann} in that although \deln $\approx 3 - 6$ appears to be 
favored by the CMB data, the standard model value of \deln = 0 is entirely 
compatible with the present data.

\section{CONCLUSIONS}

In Figure 5 we choose the priors corresponding to Case C (``$\Lambda$CDM") 
to illustrate the confrontation between the BBN constraints and those from 
the CMB. As already alluded to above, the points in the $\eta$ -- \deln 
plane (Fig.~2) are projections from a multi-dimensional parameter space 
and the relevant values of those additional parameters may not always be 
consistent with other, independent observational data.  As an illustration, 
in Figure 5 we also show three isochrones, for 11, 12, and 13 Gyr.  

\begin{figure}[ht]
	\centering
	\epsfysize=4.25truein 
\epsfbox{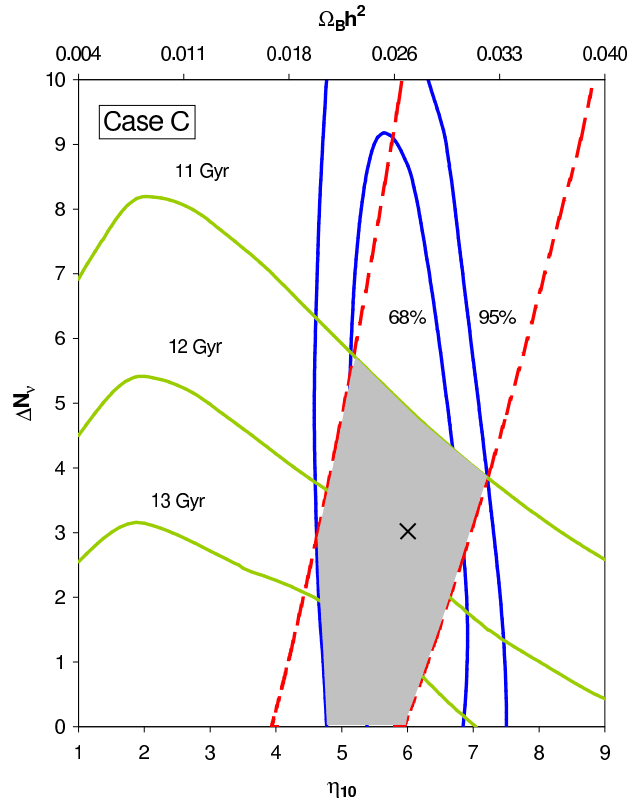}
	\caption{\small{The BBN (dashed) and CMB (solid) contours in 
	the \deln -- $\eta_{10}$ plane for the priors corresponding 
	to Case C (see Fig.~2).  The corresponding best fit isochrones 
	are shown for 11, 12, and 13 Gyr.  The shaded region delineates 
	the parameters consistent with BBN, CMB (at 95\%), and t$_{0} > 
	11$ Gyr.}}
 	\label{fig5}
\end{figure}
  
The trend in the isochrones is easy to understand: as \deln increases, 
so too do the corresponding values of the matter density ($\Omega_{M}$) 
and the Hubble parameter (H$_{0}$) which minimize $\chi^{2}$.  In addition, 
since $\Omega_{M} + \Omega_{\Lambda} = 1$, $\Omega_{\Lambda}$ decreases.  
All of these lead to younger ages for larger values of $\Delta N_\nu$. 
Note that if a constraint is imposed that the universe today is at least
11 Gyr old \cite{chaboyer}, then the BBN and CMB overlap is considerably
restricted (to the shaded region in Figure 5).  Even with this constraint 
it is clear that there is room for modest ``new physics'' (\deln $\la 6$; 
$\xi_{e} ~\la 0.3$), for which there is a limited range of baryon density 
($0.018~\la \Omega_{B}h^{2}~\la $ 0.026) which is concordant with both 
the BBN and CMB constraints.  If instead we were to impose a stricter, but 
still reasonable, constraint on the age, say that the Universe be older 
than 13 Gyr, the acceptable range of baryon density and ``new physics'' 
would be considerably narrowed.  

BBN alone does not provide any significant constraint on the magnitude 
of the ``new physics" arising from neutrino degeneracy; larger values 
of $\xi_{e}$ and \deln simply correspond to larger values of $\eta$ 
(see \cite{ks}, \cite{ostw}).  In this paper we have shown that CMB 
observations can constrain \deln (and, correspondingly, $\xi_{e}$) 
but this constraint is sensitive to the priors chosen when fitting 
the CMB data.  However, we have noted that if an additional cosmological 
constraint (on the age of the universe) is imposed, this ambiguity 
can be eliminated and a restricted range of parameters is identified: 
$\Omega_{B}h^{2} \approx 0.018 - 0.026$, \deln $\la 6$, and $\xi_{e}~\la 
0.3$.  If the extra relativistic energy density ($\Delta N_\nu$) is 
contributed by degenerate $\nu_{\mu}$, and/or $\nu_{\tau}$, then (see 
eq. 3) $\xi_{\mu}~\la 3.1$ (for $\xi_{\tau} = 0$ or, vice-versa) or, 
$\xi_{\mu} = \xi_{\tau}~\la 2.3$.

\acknowledgments
\vskip 0.2in
We thank U. Seljak and M. Zaldariagga for the use of CMBFAST 
\cite{CMBFAST} and Manoj Kaplinghat for many useful discussions.  
We thank Lloyd Knox for use of RADPACK.  This work was supported 
in part by the DOE (DE-FG02-91ER40690).

\end{document}